\documentclass[twocolumn,preprintnumbers,amssymb,amsmath,superscriptaddress,letterpaper,nofootinbib]{revtex4}[12pt]
\usepackage{times,graphics}
\usepackage[colorlinks]{hyperref}

\usepackage{graphicx}
\usepackage{dcolumn}
\usepackage{bm}
\usepackage{natbib}
\usepackage{graphicx}
\usepackage{epsfig}
\usepackage{epstopdf}
\usepackage{multirow}
\usepackage{amsmath}
\usepackage{lipsum}

\newcommand{\aj}{{Astron.~J.}}

\newcommand{\be}{\begin{equation}}
\newcommand{\ee}{\end{equation}}
\newcommand{\bea}{\begin{eqnarray}}
\newcommand{\eea}{\end{eqnarray}}

\makeatletter
\newcommand*{\shifttext}[2]{%
	\settowidth{\@tempdima}{#2}%
	\makebox[\@tempdima]{\hspace*{#1}#2}%
}
\makeatother

\newcommand{\CC}{\Lambda}

\begin{document}

\title{Running vacuum versus Holographic dark energy: a cosmographic comparison}

\author{Mehdi Rezaei}
\affiliation{Iran Meteorological Organization, Hamedan Research Center for Applied
Meteorology, Hamedan, Iran.}
\author{Joan Sol\`a Peracaula}
\affiliation{Departament de F\'\i sica Qu\`antica i Astrof\'\i sica  and   Institute of Cosmos Sciences,  Universitat de Barcelona, \\ Avinguda Diagonal 647 E-08028 Barcelona, Catalonia, Spain.}

\date{\today}

\begin{abstract}
We perform a comparative study of  different types of dynamical dark energy models (DDEs)  using the cosmographic method.  Among the models being examined herein we have  the Running Vacuum models (RVMs), which have demonstrated considerable ability to fit the overall cosmological data at  a level comparable to the standard cosmological model, $\Lambda$CDM,  and capable of alleviating  the $\sigma_8$ and $H_0$ tensions.  At the same time  we address  a variety of Holographic dark energy models (HDEs) with different options for the time (redshift)-varying model parameter $c=c(z)$. We deal with the HDEs under the double assumption of fixed and evolving holographic length scale and assess which one is better.  Both types of DDEs  (RVMs and HDEs) are confronted with the most robust cosmographic data available, namely the Pantheon sample of supernovae data (SnIa),  the baryonic acoustic oscillations data  (BAOs) extracted from measurement of the power spectrum and bispectrum  of the BOSS data release, and the cosmic chronometer measurements of the Hubble rate (CCHs)  at different redshifts   obtained from spectroscopic observations of passively evolving galaxies.  Using these data samples we assess  the viability of the mentioned DDEs   and compare them with the concordance $\Lambda$CDM model.  From our cosmographic analysis we conclude that the  RVMs  fare comparably  well to the $\Lambda$CDM, a fact which adds up more credit to their sound phenomenological status. In contrast, while some of the  HDEs are favored using  the current Hubble horizon as fixed holographic length, they become highly unfavoured in the more realistic case when the holographic length is dynamical and evolves as the  Hubble horizon.
\end{abstract}
\maketitle

\section{Introduction}

The evidence in favor of the cosmic acceleration seems nowadays well established. The primary observational source supporting such evidence is the data  on the luminosity distances of type Ia supernovae (SnIa)\citep{Riess1998,Perlmutter1999,Kowalski2008,Scolnic:2017caz}.  It  is a robust  kind of data which we categorize as cosmographic data, namely data at low redshifts which can be treated to a large extend in a model-independent way.  Fortunately,  the SnIa  is  not the only source of  cosmographic data, e.g. the imprint at low redshifts left by the  Baryon Acoustic Oscillations (BAOs)  on the large scale structure observations stemming from the primordial plasma oscillations  is also  a precious source of cosmographic observations.  These are sound wave perturbations impinging on the large scale structure (LSS) formation.  The same applies to the measurements of the Hubble rate obtained at different redshifts from the spectroscopic observations of passively evolving galaxies.  All these data are taken at relatively low redshift and hence are more accessible to us.  Many other sources of evidence  are of course available, even if they are not of cosmographic type.  In modern times, the speeding-up of the cosmic expansion is also confirmed through the precise measurements of the  anisotropies of the cosmic microwave background (CMB)\,\citep{Aghanim:2018eyx}.  Intertwining data of  different nature is of course very helpful for improving the information on the cosmological expansion history of the universe. The  data synergy makes the study  more complete and it is surely a positive aspect.  There is, however, a possible drawback in doing so in that some of the data are much more model dependent than others and the mixture may hamper our ability to disentangle the genuine effects associated to the data from the peculiar biases introduced by the different models, something that one can never avert completely but may try to mitigate as much as possible.  It is for this reason that for the present study of the cosmic acceleration we would like to focus on cosmographic data only. The cosmographic  approach proceeds by making a  minimal number of  assumptions on the expansion dynamics, namely, it does not assume any particular form of the Friedmann equations.  It only relies on the assumption that the spacetime geometry is well described on large scales by the homogeneous and isotropic  Friedmann-Lema\^\i tre-Robertson-Walker   (FLRW) metric.

 The concordance cosmological model (aka $\Lambda$CDM) postulates  the existence of a cosmological constant or $\CC$  as the ultimate cause of the cosmic acceleration, and  also the existence of large amounts of  cold dark matter (CDM) as the dominant matter component of the universe together with the necessary fraction of baryons which conform the shining stars, galaxies and other dimmer structures.  These are the fundamental ingredients of the universe, according to the concordance model. This picture, however ad hoc and imperfect as it may be,  seems to be consistent with most of the cosmological observations, but not quite.  The theoretical problems associated to the $\CC$-term\,\cite{Weinberg1989} are actually  not circumvented by  any other DE proposal\,\cite{Sola:2013gha,Sola:2013gha1,SolaPeracaula:2022hpd}.  Moreover, recently, the possibility that the vacuum energy density (VED) can be a dynamical component of the universe (and therefore that the $\CC$-term may not be that absolutely rigid quantity after all) has been put forward and linked  with fundamental physics, both within quantum field theory\,\cite{Moreno-Pulido:2020anb,Moreno-Pulido:2022phq,Moreno-Pulido:2022upl} and string theory, \cite{Mavromatos:2020kzj} as part of the potential  quantum gravity phenomenology\,\cite{QGMM2022}.  It is therefore natural to put this possibility to the test already at the level of the cosmographic data. The cosmographic test constitutes the  most basic checkout to be passed successfully by any model aspiring to describe our universe.  We perform such a test here and at the same time compare with other competing models.

This is all the more necessary if we take into account that  in the last few years some of the cosmological  observations show that the $\Lambda$-cosmology is pestered with persistent tensions concerning the prediction of some relevant cosmological parameters\,\cite{WhitePaper2022,DiValentino:2020zio,Perivolaropoulos:2022}. In particular,  the Hubble constant $H_0$ measured locally and the one extracted from the sound horizon obtained from the CMB data provide the two chief absolute scales at opposite ends of the visible expansion history of the Universe\,\citep{DiValentino:2020zio}.  By improving the accuracy of the $H_0$ measurement from Cepheid-calibrated type Ia supernovae (SnIa), evidence has been growing of a prominent discrepancy between the two. The local and direct determination gives $H_0=74.03\pm 1.42$ km/s/Mpc from \citep{Riess:2019cxk}, while the one inferred from Planck CMB data using $\Lambda$CDM cosmology  predicts $H_0 = 67.4\pm0.5$ km/s/Mpc \citep{Aghanim:2018eyx}, hence an notable tension of  $4.4 \sigma$.  The other significant (though lesser) tension concerns the discordancy among the amplitude of matter fluctuations from large scale structure (LSS) data \citep{Macaulay:2013swa}, and the value predicted by CMB experiments based on the $\Lambda$CDM, see\,\citep{DiValentino:2020vvd} for an updated status. Let us also mention that  the Baryon Acoustic Oscillations (BAOs) measured by the Lyman-$\alpha$ forest, reported in \citep{Delubac:2014aqe}, predicts a smaller value of the matter density parameter ($\Omega_{\rm m}$) compared with those obtained from CMB data.

To overcome or at least  to alleviate the problems of $\Lambda$-cosmology, different kinds of DE models have been proposed in the literature. For a summary of the many possible models, see e.g. \citep{DiValentino:2019jae,DiValentino:2021izs} and references therein. Some of the proposed models have good consistency with observational data while some others have been ruled out by cosmological  observations, see also \citep{Malekjani:2016edh,Rezaei:2017yyj, Rezaei:2017hon,Malekjani:2018qcz,Lusso:2019akb,Rezaei:2019roe,Rezaei:2019hvb,Lin:2019htv,Rezaei:2019xwo,Rezaei:2020mrj,Rezaei:2021qpq}.

In order to mitigate the $H_0$ and $\sigma_8$ tensions afflicting the $\Lambda$CDM model the idea of Dynamical Dark Energy models (DDEs) has been considered and in fact can be very helpful\,\citep{Sola:2018sjf,Gomez-Valent:2020mqn}. In our previous papers \citep{Rezaei:2019xwo,Rezaei:2021qwd}, we showed that the aforementioned subclass of the`running vacuum models' (RVMs) \citep{Sola:2013gha,SolaPeracaula:2022hpd} can be very useful, see also\,\citep{SolaPeracaula:2018xsi,Sola:2017znb} and in particular the latest fits presented in \citep{Sola:2021txs}.      For previous devoted  analyses of the RVMs and their confrontation with data, see e.g. in \citep{Gomez-Valent:2014rxa,Gomez-Valent:2014fda,Sola:2015wwa,Sola:2016jky,Sola:2017jbl,Banerjee:2019kgu}.  As indicated, the special theoretical status of the RVM's stems from the fact that these models can be derived from QFT in curved spacetime \citep{Moreno-Pulido:2020anb,Moreno-Pulido:2022phq} and from general arguments based on the renormalization group \citep{Sola:2013gha,SolaPeracaula:2022hpd,Sola:2015rra}.

The cosmographic approach as a way to distinguish between different DE scenarios was proposed in \citep{Sahni:2002fz,Alam:2003sc}.  The method has also been discussed e.g. in \citep{Visser:2003vq,Cattoen:2007sk,Capozziello:2013wha}.  We summarize it in  Sect.\,\ref{sect:cosmography} and refer the reader e.g.  to \,\cite{Rezaei:2020lfy,Rezaei:2021qwd}  for more details.  Interestingly, some cosmographic analyses lead to deviations from the concordance model, see e.g.  \citep{Capozziello:2011tj}  and \citep{Gomez-Valent:2018hwc,Gomez-Valent:2018gvm}.  In these works,  it is shown that on pure cosmographic grounds the confidence level at which we may claim that the Universe is currently accelerating is moderate,  if we rely on the data on SnIa+CCH only, whereas it is very strong  when  BAOs are also included.  Similarly, in the cosmography analysis of \citep{Capozziello:2018jya}, the tensions with the concordance model at low redshift are also confirmed.  Let us also mention the intriguing results from \citep{Lusso:2019akb}, which suggest a tension between the best fit cosmographic parameters and the $\Lambda$ cosmology at $\sim 4\sigma$ when high redshift  data points on quasars (QSO) and gamma-ray bursts (GRBs) are added to the analysis.  In previous works\,\cite{Rezaei:2020lfy,Rezaei:2021qwd} we have confirmed these results for the $\CC$CDM.  However, here we do not use these more exotic data.

Following the line of previous studies, in the current work we put  the  RVMs to the test  and verify if their success with the cosmological data\,\cite{Rezaei:2019xwo} is reinforced also within the cosmographic analysis.   At the same time we explore comparatively the HDE models with a $c^2$ term  varying with redshift and using as holographic length  both the current Hubble horizon  $L={H_0}^{-1}$ as well as the time evolving  one $L=H^{-1}(t)$.  We fit  these models to a common set of robust cosmographic data (SnIa+BAO+CCH)  and contrast them also with corresponding the yield of the concordance model.

The layout is as follows:  In Sect.\, \ref{sect:cosmography}, we briefly review the cosmographical formalism and introduce our data sets. In Sect.\,\ref{sect:hde}  we  introduce the various DDE models under consideration in this work: two characteristic types of RVM and   four different dynamical forms of the  $c^2$ parameter for the HDM models, all of them tested with fixed and variable Hubble horizon. In Sect.\,\ref{sec:numerical} we provide the results of our  numerical analysis  for the different DE models as well as a discussion about the model selection  according to the information criteria which we have used.   Finally, in Sect.\,\ref{conlusion}  we deliver our main conclusions.  An appendix at the end provides useful calculational details referred to from the main text.

\section{cosmographic approach and data sets}\label{sect:cosmography}

Cosmography is a practical approach to cosmology which helps us to obtain  information from observations mostly in a model independent way, see e.g. \citep{Aviles:2012ir, Gruber:2013wua, Capozziello:2014zda, Aviles:2014rma, Capozziello:2015rda, Aviles:2016wel, delaCruz-Dombriz:2016bqh, Capozziello:2017ddd, Capozziello:2019cav} where a variety of models is tested using this technique. There are different methods of model independent approaches which have been used in cosmology, see also \citep{Mehrabi:2021cob} for a detailed discussion and references therein.   In such an approach one can investigate the evolution of the Universe by assuming the minimal priors of isotropy and homogeneity. Adopting the Taylor expansions of the basic observables,  starting from the scale factor $a(t)$, we can obtain useful information about the cosmic flow and its evolution \citep{Demianski:2016dsa}. Assuming only the FLRW metric we can obtain the distance - redshift relations from these Taylor expansions. Therefore one could have, in principle,  fully model independent relations.
Using the first derivatives of scale factor, one can define the cosmographic functions. We limit ourselves to the following ones:

\begin{eqnarray}\label{eq1}
\text{Hubble function:}~~& H(t)=\frac{1}{a}\frac{da}{dt}\;,
\end{eqnarray}

\begin{eqnarray}\label{eq2}
\text{deceleration function:}~~& q(t)=-\frac{1}{aH^2}\frac{d^2a}{dt^2}\;,
\end{eqnarray}

\begin{eqnarray}\label{eq3}
\text{jerk function:}~~& j(t)=\frac{1}{aH^3}\frac{d^3a}{dt^3}\,,
\end{eqnarray}

\begin{eqnarray}\label{eq4}
\text{snap function:}~~& s(t)=\frac{1}{aH^4}\frac{d^4a}{dt^4}\;,
\end{eqnarray}

\begin{eqnarray}\label{eq5}
\text{lerk function:}~~& l(t)=\frac{1}{aH^5}\frac{d^5a}{dt^5}\,.
\end{eqnarray}

Although we have included for completeness the fourth and fifth derivatives of the scale factor, which define the snap and lerk functions, the errors involved in its determination are too high, as we shall see, and are therefore not sufficiently useful for discriminating between models -- cf. also  \cite{Rezaei:2020lfy,Rezaei:2021qwd} ). It means that in practice our cosmographic tools will be the first three parameters in the above list.

Setting $a=1$ in cosmographic functions we obtain the current values of the cosmographic parameters ($H_0,q_0,j_0,s_0, l_0$). Assuming these parameters and the relations between them and  $H$ derivatives, we can compute the Taylor Series expansion of the Hubble parameter, up to $z^4$ around $z = 0$.   Defining the standard  `$y$-redshift' as $y=\frac{z}{z+1}$, as introduced in \citep{Cattoen:2007sk}, one can show that the normalized Hubble rate with respect to the current value can be written in terms of the above parameters as follows, see e.g. \citep{Rezaei:2020lfy,Rezaei:2021qwd} for more details:
\begin{eqnarray}\label{ey}
E(y)\equiv\dfrac{H(y)}{H_0} \simeq 1+A y+\dfrac{B y^2}{2}+\dfrac{C y^3}{6}+\dfrac{D y^4}{24}+\cdots
\end{eqnarray}
The coefficients $A,  B, C$ and $D$  are given in terms of the current values of the above defined cosmographic parameters:
\begin{eqnarray}\label{k1}
A=1+q_0\;,
\end{eqnarray}
\begin{eqnarray}\label{k2}
B=2-q^2_0+2q_0+j_0\;,
\end{eqnarray}
\begin{eqnarray}\label{k3}
C=6+3q^3_0-3q^2_0+6q_0-4q_0j_0+3j_0-s_0\;,
\end{eqnarray}
\begin{eqnarray}\label{k4}
&&D=-15q^4_0+12q^3_0+25q^2_0j_0+7q_0s_0-4j^2_0-16q_0j_0\nonumber \\
&&\phantom{xxx}-12q^2_0+l_0-4s_0+12j_0+24q_0+24\,.
\end{eqnarray}

Using this approach, we may obtain the best fit values of the cosmographic parameters in a model independent way. Following the method which have been used in \citep{Rezaei:2020lfy,Rezaei:2021qwd}, we select the initial values of the free parameters to run the  Markov Chain Monte Carlo (MCMC) algorithm. Using this method, the free parameters being directly constrained by the cosmographic data are $(q_0, j_0, s_0, l_0)$. 
It is shown that the results of $q_0$ and $j_0$ are fully independent from the initial values of the free parameters. 
In order to guarantee that the MCMC procedure sweeps the whole of the parameters space, we assume big values of $\sigma$ for each of the free parameters. In this way  we avoid the risk of picking out local best fit values in the parameters space. In this work, the selected initial value for each of the free parameters are $q_0=-0.8, j_0=2.5, s_0=-2.0, l_0=-3.0$ (cf. Table I). In \citep{Rezaei:2020lfy,Rezaei:2021qwd}, we have examined different priors for free parameters. As one can see e.g.  in table 1 of \citep{Rezaei:2020lfy},  starting from $j_0=-5.0$ or $j_0=5.0$, MCMC leads to $j_0>0$ up to more than $5\sigma$.  Therefore, we find that the $j_0$ values vary  in a positive range (i.e. we always  have $j_0>0$). Thus,  in order to save time for running the MCMC algorithm, we have set a positive range for the  $j_0$ prior.

In order to put constrains on the cosmographic parameter using low redshift observational data, we set the value of cosmographic parameters ($q_0,j_0,s_0$ and $l_0$) as the free parameters in MCMC algorithm. Then, the best fit values for these parameters are those which can minimize the $\chi^2$ function. The low-redshift observational data sets which we have used in this work are as follows:

\begin{itemize}
\item  SnIa. We have used the Pantheon data sample of supernovae.  It involves  the set of the latest 1048 data points for the apparent magnitude of type Ia SnIa within the range $0.01 < z < 2.26$\, \citep{Scolnic:2017caz} ;

\item BAOs.  Insofar as  the baryonic acoustic oscillations data are concerned we have used the radial component of the anisotropic BAOs extracted from measurements of the power spectrum and bispectrum from the BOSS,  Data Release 12,  galaxy sample \citep{Gil-Marin:2016wya}, the complete SDSS III $Ly\alpha$-quasar \citep{duMasdesBourboux:2017mrl} and the SDSS-IV extended BOSS  DR14 quasar sample in the relatively low redshift range $0.8<z<2.2$, see Ref. \citep{Gil-Marin:2018cgo}.

\item  CCHs.  These are the so-called cosmic chronometer data, namely  data points on the Hubble rate,  $H(z)$,  obtained from spectroscopic  techniques applied to passively–evolving galaxies, i.e.  galaxies with old stellar populations and low star formation rates. The CCH  data constitute an ideal tool  for obtaining observational values of the Hubble function within the redshift range $z<2$ derived from galaxies located at different angles in the sky. Most importantly, these data are uncorrelated with the BAO data points.  We have used as much as  31 data points of this type in the present work from different references \citep{Zhang:2012mp,Jimenez:2003iv,Simon:2004tf,Moresco:2012jh,Moresco:2016mzx,Ratsimbazafy:2017vga,Stern:2009ep,Moresco:2015cya}.  By virtue  of the Cosmological Principle, the dependence of these data points on the angle and location of the measured galaxies is removed. Therefore  CCHs are just functions of the redshift. These measurements are also independent of the Cepheid distance scale and do not rely on any particular cosmological model.  We point out  that the CCHs data were not used in our previous comparative study \cite{Rezaei:2021qwd}, in which, in contrast, higher redshift data on QSOs and GRBs were employed.  

\end{itemize}


\section{Cosmological dark energy models}\label{sect:hde}

In this section we introduce the dynamical dark energy models (DDEs) which will be subject of cosmographic analysis in this work.  We start with the class of running vacuum models (RVMs) and subsequently we present a variety of holographic dark energy models (HDEs) and compare them all among themselves and also with the concordance $\CC$CDM model.

\subsection{Running vacuum models}
The RVMs, which have proven to be perfectly consistent from the phenomenological point of view \citep{Sola:2021txs},  have acquired a especial significance nowadays since  their theoretical structure emerges  from the renormalization of the effective action  of  QFT in curved spacetime, as has been recently shown in the comprehensive works \citep{Moreno-Pulido:2020anb,Moreno-Pulido:2022phq},  see also  \cite{SolaPeracaula:2022hpd} for a review.  It is also remarkable that their structure can also be motivated  within the low energy effective action of string theory, see\,\citep{Mavromatos:2020kzj}.
The effective form of the vacuum energy density for the RVM models can be presented as follows\,\citep{Sola:2015rra}:
\begin{equation}\label{rLRVM}
\rho_{\rm RVM}(H) = \frac{\Lambda(H)}{\kappa^2}=
\frac{3}{\kappa^2}\left(C_0 + \nu H^{2} +\tilde{\nu} \dot{H} + \alpha
\frac{H^{4}}{H_{I}^{2}} + \dots \right) \,,
\end{equation}
Dots stand for higher order powers of $H$;   $\kappa^2=8\pi G$,  and $C_0 >0$ is a constant closely related to the (current era) cosmological constant. The coefficients $\nu$, $\mu$ and $\alpha$ are small since they control the (mild) dynamics of the vacuum energy density.  When these coefficients vanish, the above expression boils down to the value of the constant cosmological vacuum energy density: $\rho_{\Lambda}=\frac{3}{\kappa^2}\,C_0$, in which the cosmological constant is precisely $\Lambda=3C_0$.   The parameter $H_I$ denotes the Hubble scale at the inflationary epoch, although it will not be used here since we analyze the current universe, so we can set $\alpha=0$ in our study. However,  in the context of QFT in curved spacetime,  explicit calculations of quantum effects lead to $\nu\neq 0$ \,\cite{Moreno-Pulido:2020anb,Moreno-Pulido:2022phq} and hence a fundamental dynamical term of ${\cal O}(H^2)$ is secured.  
Since $0<|\nu|<1$, the correction is of course small and the RVM behaves approximately as the $\CC$CDM. Still, the presence of $\nu\neq 0$ can be significant and in fact it can help sorting out the tensions afflicting the $\CC$CDM, see e.g. \citep{Rezaei:2019xwo,Sola:2021txs} and previous works such as \citep{Sola:2016jky,Sola:2017jbl}.

The RVM equation of state is that of an ideal de-Sitter fluid, despite the time dependence of the vacuum energy, therefore
$p_{\rm RVM} (H) = - \rho_{\rm RVM} (H)$, where $p_{\rm RVM} (H)$ denotes the vacuum pressure density.  As indicated before, around the current epoch and for all the practical cosmographical considerations relevant to the present study, it is sufficient to approximate the expression \eqref{rLRVM} by ignoring all the terms ${\cal O}(H^4)$. The latter, however, are crucial to account for inflation in the early universe within the context of the RVM\,\citep{Lima:2012mu,Perico:2013mna,Sola:2015rra,Sola:2015csa}. The coefficient  $\alpha$ will therefore play no role here  for the aforementioned reasons. However,  coefficients $\nu$ and  $\tilde{\nu}$  will be sensitive to our fits, as we shall see in the course of our analysis.

In previous works \citep{Rezaei:2019xwo,Rezaei:2021qwd}  we investigated two scenarios for DDE whose energy density can be expressed as a power series expansion of the Hubble rate (and its time derivatives).  The DDE models being studied in that work are:  i) ghost DE models, and   ii)  `running vacuum models' (RVMs) as given in Eq.\,\eqref{rLRVM}.  The ghost DE models were not favored by our  analyses. The RVMs, in contrast, were significantly supported.  
Quite obviously, therefore, the RVMs   are theoretically  well-motivated.   At the present epoch, the relevant terms of the power series expansion \citep{Sola:2017znb}. can only be of order $H^2$ at most (this includes $\dot{H}$), whereas the higher orders  $H^n (n>2)$ can be used in the early Universe to successfully implement inflation, see e.g.\,\citep{Lima:2012mu,Perico:2013mna,Sola:2015rra,Sola:2015csa}.

More specifically, the two running models models addressed in this study read as follows:
\begin{eqnarray}\label{gde3}
R1: \rho_{\rm de}(z)&=&\dfrac{3}{\kappa^2}\left[ C_0+\nu H^2(z)\right] \,,\\
R2: \rho_{\rm de}(z)&=&\dfrac{3}{\kappa^2}\left[C_0+\dfrac{2}{3}\mu \dot{H}(z)+\nu H^2(z)\right]\label{gde6}\,.
\end{eqnarray}
Obviously R1 is  a particular case of R2 (with $\mu=0$), but we keep it because it represents the canonical realization of the RVM \,\citep{Sola:2013gha,SolaPeracaula:2022hpd,Sola:2015rra} and has one parameter less.  These models are also particular cases of the more general RVM structure \eqref{rLRVM}. We have redefined the parameter $\tilde{\nu}$ of \eqref{rLRVM}  as  $\tilde{\nu}\equiv\dfrac{2}{3}\mu$ just for convenience and also for better comparison with our previous work \citep{Rezaei:2019xwo}.
In the above equations, the  parameter $C_0$ has dimension $+2$ in natural units, see below.

It is important to emphasize that the vacuum energy density of the  RVMs  includes  a nonvanishing (constant) additive term ($C_0\neq 0)$, as indicated in Eq.\,\eqref{rLRVM}\,\citep{Sola:2013gha,SolaPeracaula:2022hpd,Sola:2015rra}.  Thus,  for $\nu,\mu\to 0$ the two RVM models  \eqref{gde3} reduce to the $\CC$CDM case. In the absence of such nonvanishing additive constant the model could not  have a smooth limit connection with the  $\CC$CDM.  This was the basic reason for the failure of the ghost dark energy models which we studied in \cite{Rezaei:2021qwd}, as they were unable to properly fit the  cosmological observations.

The above two models R1 and R2  can be realized both as strict vacuum models or with a nontrivial EoS slightly departing from $-1$, which is actually possible for the quantum vacuum owing to the presence of quantum effects renormalizing the classical result\,\cite{Moreno-Pulido:2022phq,Moreno-Pulido:2022upl}. We choose the dynamical EoS option here and we implement this possibility by assuming that the RVM density is covariantly  self-conserved:
$\dot{\rho}_{de} + 3H(\rho_{de}+p_{de}) = 0$,
%
independently of the equation of matter and radiation  self-conservation.  Therefore, for the current work we adopt for the RVMs the same local conservation laws as  for the HDE models studied in the next section (see also the appendix for more details).  In this sense the comparison with the two sorts of models can be made more balanced.  Let us point out that the same RVMs can be studied under the assumption of interaction with matter, see \citep{Sola:2016zeg} for a review, or even under the assumption of time-varying gravitational coupling. In all these contexts the Bianchi identity is satisfied and the vacuum energy density becomes dynamical throughout the cosmic expansion\,\cite{FritzschSola}.   In our case the departure from $-1$ is mimicked phenomenologically by the fact that we can allow for matter and  DE to be both self-conserved and this condition implies a nontrivial EoS for the vacuum, as previously noted. This same setting was the one we adopted in our previous study\,\citep{Rezaei:2019xwo}  (see also \citep{Gomez-Valent:2015pia}) in which we performed a non-cosmographic analysis involving  CMB  and structure formation data (LSS).

The calculation of the cosmographic parameters for the RVMs follows the same lines as in \citep{Rezaei:2019xwo} and we need not repeat it here. We limit ourselves to quote the result for Model R2.  For small redshift $z$, i.e. around our current epoch where the cosmographic analysis holds,  the effective EoS for such model reads
\begin{eqnarray}
\label{eq:EoSmod3}
w_{\rm de}(z)&\simeq&-1+\frac{H_0^2(1-\nu)}{C_0}\,\Omega_{m,0}\,(\nu-\mu)\,(1+z)^3\nonumber\\
&\simeq& -1+\frac{\Omega_{m,0}}{1-\Omega_{m,0}}\,(\nu-\mu)\,(1+z)^3\,,
\end{eqnarray}
Here  $C_0$ is the same coefficient of Eq.\,\eqref{gde3}. It can be obtained after matching the current value of the vacuum energy density   to the value of the measured density associated to the cosmological constant. One finds, taking only  linear order in $\nu$ and $\mu$:
\begin{equation}\label{35}
C_0=H^2_0[\Omega_{\rm de,0}-\nu+\mu(1+w_{\rm de,0}\Omega_{\rm de,0}+\dfrac{\Omega_{\rm r,0}}{3})]\;.
\end{equation}
The corresponding EoS for model R1 is just obtained from the previous formulas by setting $\mu=0$. We can see that for $\mu,\nu\to 0$ we recover the EoS of the $\CC$CDM, i.e. $w_{\rm de}\to -1$.
The above EoS  shows that the RVM's behave effectively as quintessence  for $\nu-\mu>0$ or phantom-like behavior for $\nu-\mu>0$.  It demonstrates that the  running vacuum models can mimic exotic behaviors  (especially in the  phantom DE case) which in the context of elementary scalar fields could be problematic.  The RVMs are perfectly well behaved and consistent with QFT since there is no real phantom at all\,\citep{Moreno-Pulido:2020anb,Moreno-Pulido:2022phq}.

\begin{table*}
 \centering
 \caption{The prior and maximum value of marginalized posterior for each of cosmography parameters and their $1\sigma$ uncertainties obtained in the model independent (MI)  approach using Pantheon+BAOs+CCH as data combination.
}
\begin{tabular}{c  c  c c c}
\hline \hline
 & $q_0$ & $j_0$ & $s_0$& $l_0$ \\
\hline
Prior we have used  & $-2.0,2.0$ &$0.0,5.0$ &$-5.0,5.0$ &$-5.0,5.0$\\
\hline
Maximum value of marginalized posterior & $-0.648\pm 0.073$ &$1.074^{+0.099}_{-0.103}$ &$-3.39^{+0.78}_{-1.04}$ &$-4.6^{+2.23}_{-1.87}$\\
\hline \hline
\end{tabular}\label{tab:bestmi}
\end{table*}

\begin{table*}
\centering
\caption{The maximum value of marginalized posterior for each of the parameters of $\Lambda$CDM model (left part) and the value of cosmographic parameters calculated upon maximum of the marginalized posterior (right part).
 }
\begin{tabular}{c c  c c c c  c}
\hline \hline
  maximum value & of marginalized & posterior  & $\vert$ & & computed values &   \\
$\Omega_{dm,0}$  & $\Omega_{b,0}$ & $h$  & $\vert$ & $q_0$&  &  $j_0$\\
\hline
$0.242\pm 0.011$ & $0.045\pm 0.007$ & $0.695\pm 0.029$ & $\vert$ & $-0.571\pm 0.018$ &  & $1.0$\\
\hline \hline
\end{tabular}\label{bestlcdm}
\end{table*}

\begin{table*}
\centering
\caption{The maximum value of marginalized posterior for each of the parameters of RVM models (left part) and the value of cosmographic parameters calculated upon maximum of the marginalized posterior (right part). We note that in model R2 we have set $\mu=-\nu$.
 }
 \begin{tabular}{c  c  c c c c c c c}
 \hline \hline
   & & maximum value & of marginalized & posterior  & $\vert$ & & computed values &  \\
 Model & $\Omega_{dm,0}$ & $\Omega_{b,0}$ & $h$ & $\nu $ & $\vert$ & $q_0$&    & $j_0$\\
 \hline
  R1& $0.250^{+0.034}_{-0.032}$ & $0.04 \pm 0.004$ & $0.695^{+0.013}_{-0.016}$ & $0.431^{+0.031}_{-0.027}$ &  $\vert$ & $-0.708^{+0.074}_{-0.093}$ &    & $1.104^{+0.131}_{-0.112}$\\
 \hline
 R2 & $0.243 \pm 0.026$ & $0.044 \pm 0.004$& $0.694^{+0.018}_{-0.011}$ & $-0.532^{+0.024}_{-0.029}$ & $\vert$ & $-0.598^{+0.109}_{-0.102}$ &   & $1.097^{+0.242}_{-0.250}$\\
 \hline \hline
\end{tabular}\label{bestdde3}
\end{table*}

\begin{table*}
 \centering
 \caption{The maximum value of marginalized posterior for each of the parameters of HDE models (left part) and the value of cosmographic parameters calculated upon maximum of the marginalized posterior (right part). We note that in these models we have $L=H_0^{-1}$ as IR cutoff.
}
\begin{tabular}{c c c  c c c  c  c c}
\hline \hline
 &  & maximum & value of marginalized & posterior & $\vert$ & & computed values & \\
Model & $\Omega_{dm,0}$ & $\Omega_{b,0}$ & $h$ & $c_1$ & $\vert$ & $q_0$&    & $j_0$\\
\hline
H01 & $0.247 \pm 0.027$ & $0.044^{+0.007}_{-0.011}$ & $0.719 \pm 0.031$ & $-0.084^{+0.011}_{-0.013}$ &  $\vert$ & $-0.580^{+0.154}_{-0.123}$ &    & $1.008^{+0.201}_{-0.214}$\\
\hline
H02 & $0.235 \pm 0.044$ & $0.046^{+0.005}_{-0.009}$& $0.721^{+0.049}_{-0.045}$ & $-0.094^{+0.014}_{-0.010}$ & $\vert$ & $-0.606^{+0.223}_{-0.229}$ &   & $1.080^{+0.370}_{-0.286}$\\
\hline
H03 & $0.261^{+0.047}_{-0.049}$ & $0.040^{+0.007}_{-0.009}$ & $0.721 \pm 0.042$ & $0.158^{+0.009}_{-0.012}$ & $\vert$ & $-0.633^{+0.104}_{-0.117}$ &  & $1.104^{+0.131}_{-0.112}$\\
\hline
H04 & $0.244 \pm 0.048$ & $0.042^{+0.007}_{-0.004}$ & $0.719^{+0.040}_{-0.043}$ & $0.410^{+0.009}_{-0.010}$ & $\vert$ & $-0.596^{+0.117}_{-0.110}$ &  & $1.091^{+0.234}_{-0.242}$\\
\hline \hline
\end{tabular}\label{besth0}
\end{table*}

\begin{table*}
 \centering
 \caption{The maximum value of marginalized posterior for each of the parameters of HDE models (left part) and the value of cosmographic parameters calculated upon maximum of the marginalized posterior (right part). We note that in these models we have $L=H^{-1}$ as IR cutoff.
}
\begin{tabular}{c c c  c c c  c  c c}
\hline \hline
 &  & maximum & value of marginalized & posterior  & $\vert$ & & computed values & \\
Model & $\Omega_{dm,0}$ & $\Omega_{b,0}$ & $h$ & $c_1$ & $\vert$ & $q_0$&    & $j_0$\\
\hline
H1 & $0.162 \pm 0.039$ & $0.024^{+0.007}_{-0.009}$ & $0.738 \pm0.029$ & $-0.371^{+0.014}_{-0.012}$ &  $\vert$ & $-0.941^{+0.027}_{-0.028}$ &    & $3.564^{+0.449}_{-0.440}$\\
\hline
H2 & $0.154^{+0.051}_{-0.049}$ & $0.038^{+0.011}_{-0.027}$& $0.698^{+0.054}_{-0.051}$ & $-0.216^{+0.021}_{-0.032}$ & $\vert$ & $-0.937 \pm 0.034$ &   & $2.984^{+0.252}_{-0.313}$\\
\hline
H3 & $0.151^{+0.044}_{-0.039}$ & $0.059 \pm 0.008$ & $0.737^{+0.047}_{-0.045}$ & $0.469^{+0.014}_{-0.012}$ & $\vert$ & $-0.908^{+0.022}_{-0.023}$ &  & $3.565^{+0.279}_{-0.273}$\\
\hline
H4 & $0.141^{+0.041}_{-0.039}$ & $0.029^{+0.007}_{-0.009}$ & $0.690^{+0.022}_{-0.026}$ & $0.638^{+0.027}_{-0.034}$ & $\vert$ & $-0.177 \pm 0.138$ &  & $0.313^{+0.154}_{-0.135}$\\
\hline \hline
\end{tabular}\label{besth}
\end{table*}

\subsection{Holographic DE models with varying  \lowercase{c}$^2$ term}

Some alternative dynamical frameworks mimicking DE are those  based on holography, which go under the name of  holographic DE models (HDEMs).  The first suggestions implementing these ideas in the cosmological context were put forward in \citep{Li:2004rb,Hsu:2004ri}.  The holographic principle, being a fundamental principle in quantum gravity, was further expanded in different directions\citep[for a detailed discussion, see][]{tHooft1993,Susskind1995,Bekenstein:1973ur,Bekenstein:1974ax,Cohen1999,Telali:2021jju,Drepanou:2021jiv}.  For a comprehensive review, see e.g. \citep{Bousso:2002ju} and references therein.
In cosmological contexts, assuming the whole of the universe, the vacuum energy related to the holographic principle has been conceived in some contexts  as holographic dark energy (HDE). However, we will consider here that the vacuum models and HDE are different approaches to the DE, and for this reason we will analyze them separately. In the HDE case,  the  DE energy density is generically given by:
\begin{eqnarray}\label{hde}
\rho_{\rm d}=3 c^2M_{\rm p}^2L^{-2}\;,
\end{eqnarray}
where  $M_{\rm p}=1/\kappa$  is the reduced Planck Mass and  $c^2$ is a dimensionless numerical parameter (or function of some cosmological variable). Finally,  $L$ is the holographic length scale,  which acts as infrared (IR) cutoff. It can  be defined in alternative ways for different HDE models. Typical proposals in the literature for  $L$ are: the Hubble length, $L = H^{-1}$, the particle horizon, the future event horizon, the (inverse square root) of the Ricci scalar $R$ and others\citep{Li:2004rb,Hsu:2004ri,Huang:2004ai,Pavon2005,Gao:2007ep,Zimdahl2007,Duran:2010ky,Radicella:2010vf,Sheykhi:2011cn}.  For constant $c^2$ it was shown by M. Li  in \cite{Li:2004rb} that within the free theory the holographic scale $L$  cannot be taken as the  Hubble length nor the particle horizon since in the first case it leads to a wrong equation of state, namely that for dust, while in the second case it does not lead to the current accelerated phase of expansion.  However, the presence of an interaction of the HDE with a pressureless component (dust matter or cold dark matter) can avoid the trouble and moreover offers a possible solution to the cosmic coincidence problem\citep{Pavon2005}.  Alternatively, if a slowly varying $c^2$ term is admitted the choice  $L = H^{-1}$ for the holographic length is once more admissible even if there are no interaction between the holographic dark energy and dust matter, thereby leading to the current accelerated expansion of the universe.  We shall explore this option here in different ways. In fact there are no strong evidence supporting that the  $c^2$ parameter in Eq.(\ref{hde}) must be constant. Adopting a time (redshift) slowly varying form of $c^2$, we can reconsider the $H^{-1}$  option as a valid holographic length scale $L$ within this new context of variable $c$. Furthermore, as another choice we can assume the current value of the Hubble parameter as the holographic length scale $L={H_0}^{-1}$.  While the second choice may not be more natural than the first, it can be quite useful to assess the sensitivity of the analysis to making the holographic length rigid rather than dynamical.
 The results of different studies indicate that the HDE models with slowly varying $c^2$ with $L={H_0}^{-1}$ or $L={H}^{-1}$ can simultaneously drive current accelerated expansion and also solve the coincidence problem \citep[see][]{Pavon2005,Pavon:2006qm,Guberina:2006qh,Radicella:2010vf}.  Particularly, in \citep{Malekjani:2018qcz} the option $L={H_0}^{-1}$ for  HDE was confronted with the latest cosmological observations,  including  data on SnIa, BAO, CMB shift parameter, BBN and expansion Hubble data, and also at the perturbations level using the latest growth rate ($f(z)\sigma_8$) data. The results of this study showed that this type of HDE models are well fitted to both expansion and growth rate observations at a comparable level  to the $\Lambda$CDM model. However, the possibility of a dynamical $L={H}^{-1}$ for the set of models explored was not addressed. Here we compare these two options, which lead to significant differences, and at the same time we contrast them with the cosmographic performance of the RVMs and the concordance $\CC$CDM model.

In what follows we concentrate on a variety of popular HDE models existing in the literature, which we would  like to test on cosmographical grounds.  Assuming the Hubble horizon as the holographic IR cutoff in the two versions $L=H^{-1}$ and $L={H_0}^{-1}$   we study the background evolution of the universe in the presence of HDE using four different parameterizations for $c^2(z)$ defining a slowly varying $c^2$ term in Eq.\,\eqref{hde}.  These parameterizations are referred to as Models H1,H2,H3 and H4 when $L=H^{-1}$, and as H01, H02, H03 and H04 when  $L=H_0^{-1}$ is chosen.  The situation with dynamical horizon had not been studied in the literature before (to the best of our knowledge), whereas the case with fixed horizon was studied previously in \citep{Malekjani:2018qcz}. 
They are defined as follows:
\begin{itemize}
\item Chevalier-Polarski-Linder (CPL) type of  parametrization \citep{Chevallier2001}:
\begin{eqnarray}\label{par1}
{\rm H1:}~~& c(z)=c_0+c_1\frac{z}{1+z}\;.
\end{eqnarray}
The parameter  $c_0$ in this equation should not be confused with $C_0$ from the RVMs, see Eq.\,\eqref{rLRVM}, even though both are of course related to the value of the cosmological constant at present.
\item Jassal-Bagla-Padmanabhan (JBP) parametrization \citep{Jassal:2004ej},
\begin{eqnarray}\label{par2}
{\rm H2:}~~& c(z)=c_0+c_1\frac{z}{(1+z)^2}\;.
\end{eqnarray}
\item Wetterich parametrization \citep{Wetterich:2004pv}:
\begin{eqnarray}\label{par3}
{\rm H3:}~~& c(z)=\frac{c_0}{1+c_1\ln (1+z)}\;.
\end{eqnarray}
\item Ma-Zhang parametrization \citep{Ma:2011nc},
\begin{eqnarray}\label{par4}
{\rm H4:}~~& c(z)=c_0+c_1\left(  \frac{\ln (2+z)}{1+z}-\ln 2\right) \;.
\end{eqnarray}
\end{itemize}
Setting $c_1=0$ in each of the above equations, we retrieve the usual HDE model with constant $c$ term, which would lead to a nonviable situation, as we have warned.   Since the dynamics of the DE must be small, we expect $|c_1|\ll c_0$ in all cases, although $c_1$ can have either sign and it can only be known after fitting the models to the data.
In a homogeneous, isotropic and  spatially flat FLRW cosmology, in which the cosmic fluid contains radiation, non-relativistic matter and dark energy, the first Friedmann equation can be written as follows:
\begin{eqnarray}\label{frid1}
H^2=\frac{1}{3M^2_{\rm p}}(\rho_{\rm r}+\rho_{\rm m}+\rho_{\rm d})\;,
\end{eqnarray}
where as before   $M_{\rm p}=1/\kappa$ is the reduced Planck Mass.  The basic functions involved are the Hubble parameter $H\equiv {\dot a}/a$, the energy density of radiation, $\rho_{\rm r}$, the energy density of pressureless matter, $\rho_{\rm m}$, and the energy density of DE, $\rho_{\rm d}$. The dimensionless density parameters $\Omega_{\rm i}$, for  the different components of the universe  can then be expressed as follows:
\begin{eqnarray}\label{omega}
\Omega_{\rm r}=\frac{\rho_{\rm r}}{3M^2_{\rm p}H^2}\,, ~ \Omega_{\rm m}=\frac{\rho_{\rm m}}{3M^2_{\rm p}H^2}\,, ~\Omega_{\rm d}=\frac{\rho_{\rm d}}{3M^2_{\rm p}H^2}\,.
\end{eqnarray}
The parameter $c_0$ in the above HDE models is obviously positive and if we use the approximation $|c_1|\ll c_0$   it is easy to see from equations \eqref{hde} and the last of \eqref{omega} that
\begin{equation}\label{eq:codetermHDE}
c_0^2\simeq \frac{\rho_{\rm d}}{3M^2_{\rm p}H^2}\simeq \frac{\rho_{\Lambda}}{3M^2_{\rm p}H^2}\simeq \Omega_\Lambda\,,
\end{equation}
where $\Omega_\Lambda\simeq 0.7$ is the approximate value of the cosmological parameter associated to the cosmological constant, a measured quantity. Therefore, as an order of magnitude estimate,  we have
\begin{equation}\label{coOmegaLambda}
c_0\simeq \sqrt{\Omega_\Lambda}\sim 0.8\,.
\end{equation}
The parameter $c_0$ is therefore not a free parameter, but its  final value  will depend, of course, on the particular model and will receive corrections of order $|c_1|/c_0\ll1$.
Assuming the two different choices $L=H^{-1}$ and $ L=H_0^{-1}$ for the effective  IR cutoff in Eq.\eqref{hde}, the cosmological equations and corresponding equations of state for each model can be found explicitly in the appendix.

\section{Numerical analysis and discussion}\label{sec:numerical}

In this section we present and discuss our numerical results concerning  the various DDEs,  including the concordance $\Lambda$ cosmology.

\subsection{Comparison with the MI approach}

For the selected data combination, we perform our analysis aimed at finding the maximum value of the marginalized posterior of the free parameters and their $1\sigma$ and $2\sigma$ uncertainties. We present the selected priors which we have used and the results of the model independent approach in Table \ref{tab:bestmi}. We observe that both the deceleration and jerk parameters are tightly constrained.  Comparing these results with those we obtained in \citep{Rezaei:2020lfy,Rezaei:2021qwd} using different data samples, we learn that the current data combination based on the robust data set SnIa+BAO+CCH lead indeed to a better constrain on the jerk parameter in the present case.   However, using the same data combination, Table \ref{tab:bestmi} shows that we can not put tight constraints on the remained cosmographic parameters \eqref{eq4} and \eqref{eq5}, namely the snap and lerk  ($s_0$ and $ l_0$), as anticipated. The results of these two parameter, strongly depend on the initial value we have used in MCMC analysis. For  ($s_0$ and $ l_0$)  we obtain larger values of the uncertainties, which means that we cannot determine them with enough accuracy. The root of the difficulty here lies on the fact that  $s_0$ and $l_0$ are involved in the  third and forth order powers of the redshift in the Taylor expansion of $H$,  see equations \eqref{k3}-\eqref{k4}.  For this reason our analysis of model independent (MI) determination shall  henceforth focus on just  $q_0$ and $j_0$.

Using our data combination and running the MCMC algorithm (along the lines of \citep{Rezaei:2021qwd}), we find the best fit values of the free parameters of the different RVM and  HDE models, including $\Lambda$CDM and corresponding  $1\sigma$ uncertainty for each of the free parameters  for each cosmology.  Using the formulas of the appendix (particularly \eqref{eq:computedq0} and \eqref{eq:computedj0} we calculate the  value of the cosmographic parameters for each model and compare it with those previously obtained in the model-independent (MI)  approach (presented in Table \ref{tab:bestmi}).
To compute the value of the cosmographic parameters  for each of the DE models, we must find the maximum value of the marginalized posterior of the model parameters and also the associated confidence regions. For the RVMs we have $\Omega_{dm,0}, \Omega_{b,0}, h, \nu$ and $\mu$ as the free parameters.  However, for model R2 (which has one more parameter as compared to R1)  we introduce the setting  $\mu=-\nu$ (reducing in one unit the number of parameters) on the same grounds as  explained in Refs. \citep{Rezaei:2019xwo,Rezaei:2021qwd,Gomez-Valent:2015pia}. For  the HDEs, instead, we have $\Omega_{dm,0}, \Omega_{b,0},h$ and $c_1$.  As for the $\Lambda$CDM case, we have $\Omega_{dm,0}, \Omega_{b,0}$ and $h$.  Proceeding in this way the number of extra free parameters  for all these DE models with respect to the $\CC$CDM  is just $1$, and therefore they are all leveled in this sense.  Use of appropriate information criteria (cf. Sec. \ref{sec:InformationCriteria}) will duly penalize these models for the existing of such an extra degree of freedom in the fit. Next, using the mentioned SnIa+BAO+CCH data combination, we find the maximum value of marginalized posterior of the model parameters for the different DE cosmologies, as well as their confidence regions at  $1\sigma$ level. We report our results in the left part of Tables (\eqref{bestlcdm}-\eqref{besth}) for the different DE scenarios. Inserting these results in Eqs. (\eqref{q}, \eqref{j}) of the appendix we may calculate the cosmographic parameter values for each DE model under consideration. The calculated results for each model, and their $1\sigma$ confidence regions, are reported in the right part of Tables (\ref{bestlcdm}-\ref{besth}). In order to obtain the  $1\sigma$ confidence regions for cosmographic parameters we have runned a MCMC algorithm in which, each of the model parameters can vary around its best fit value up to  $1\sigma$ confidence level. In each of the chains, we obtain a unique value for $q_0$  and $j_0$.

\begin{figure*}
	\centering
	\includegraphics[width=15cm]{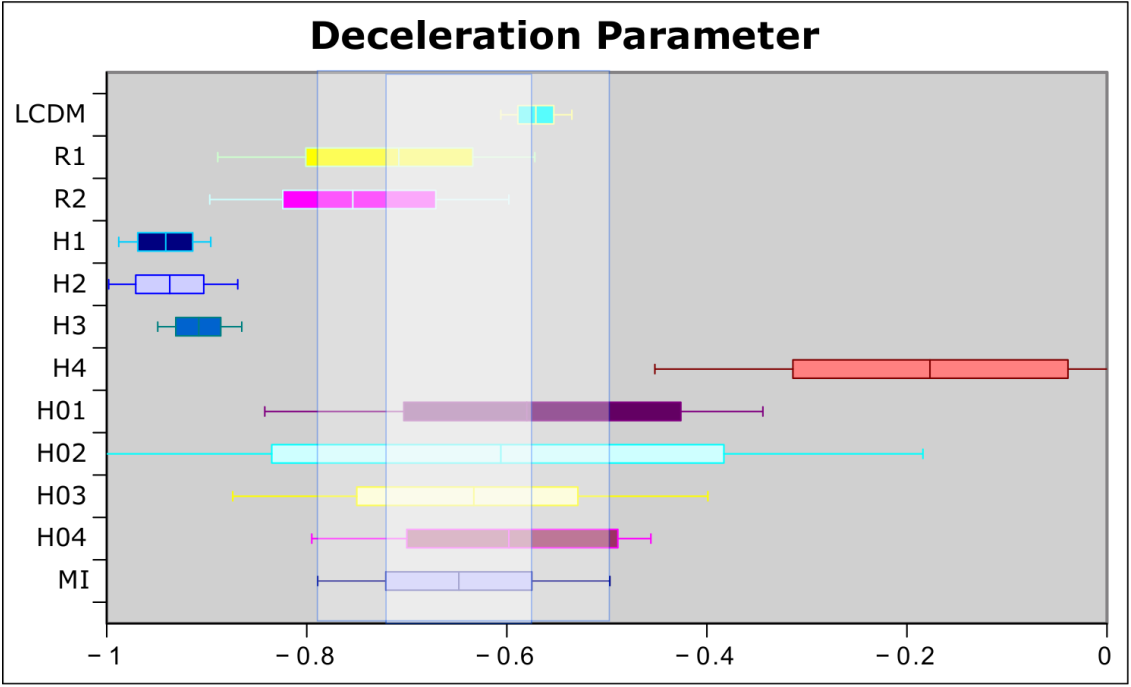}
		\caption{The computed values of $q_0$  for the different DE models,  using the best fit value of the model parameters and Eq.\eqref{eq:computedq0} in the appendix for $z=0$, as compared to the model-independent (MI) approach.}
	\label{fig:q}
\end{figure*}

\begin{figure*}
	\centering
	\includegraphics[width=15cm]{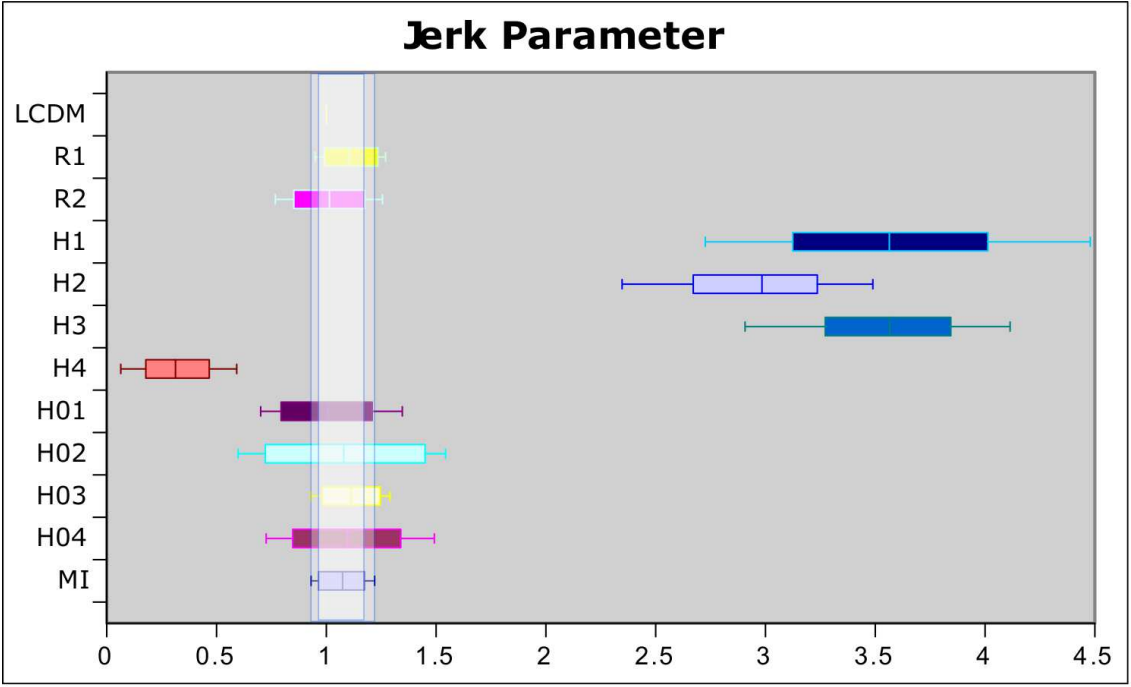}
		\caption{Same as in Fig.\,\eqref{fig:q}, but for the computed values of $j_0$ for the different DE scenarios. Use is made of equations \eqref{eq:computedq0} and \eqref{eq:computedj0} in the appendix for $z=0$}.
	\label{fig:j}
\end{figure*}

\begin{figure*}
	\centering
	\includegraphics[width=8.5cm]{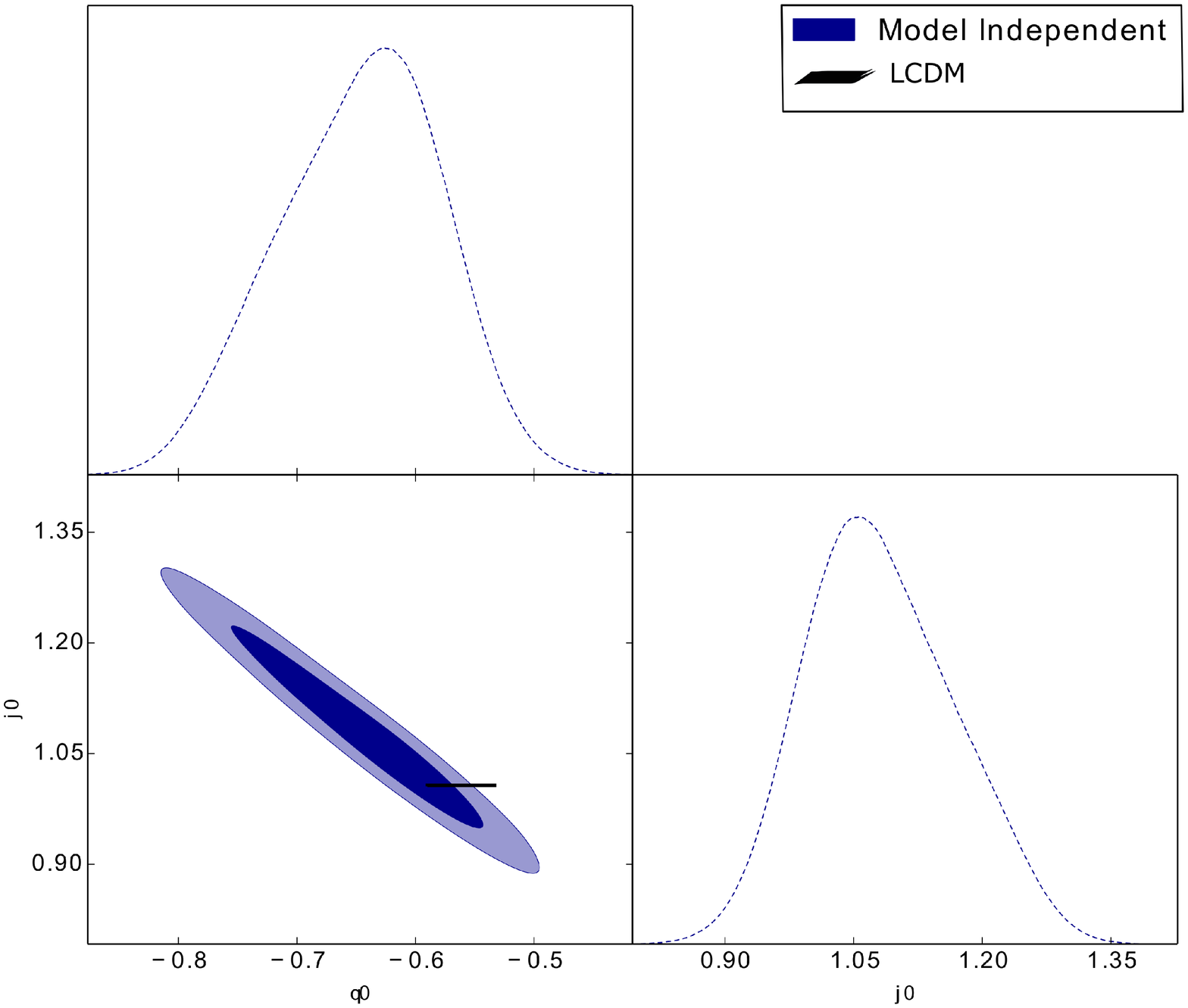}
	\includegraphics[width=8.5cm]{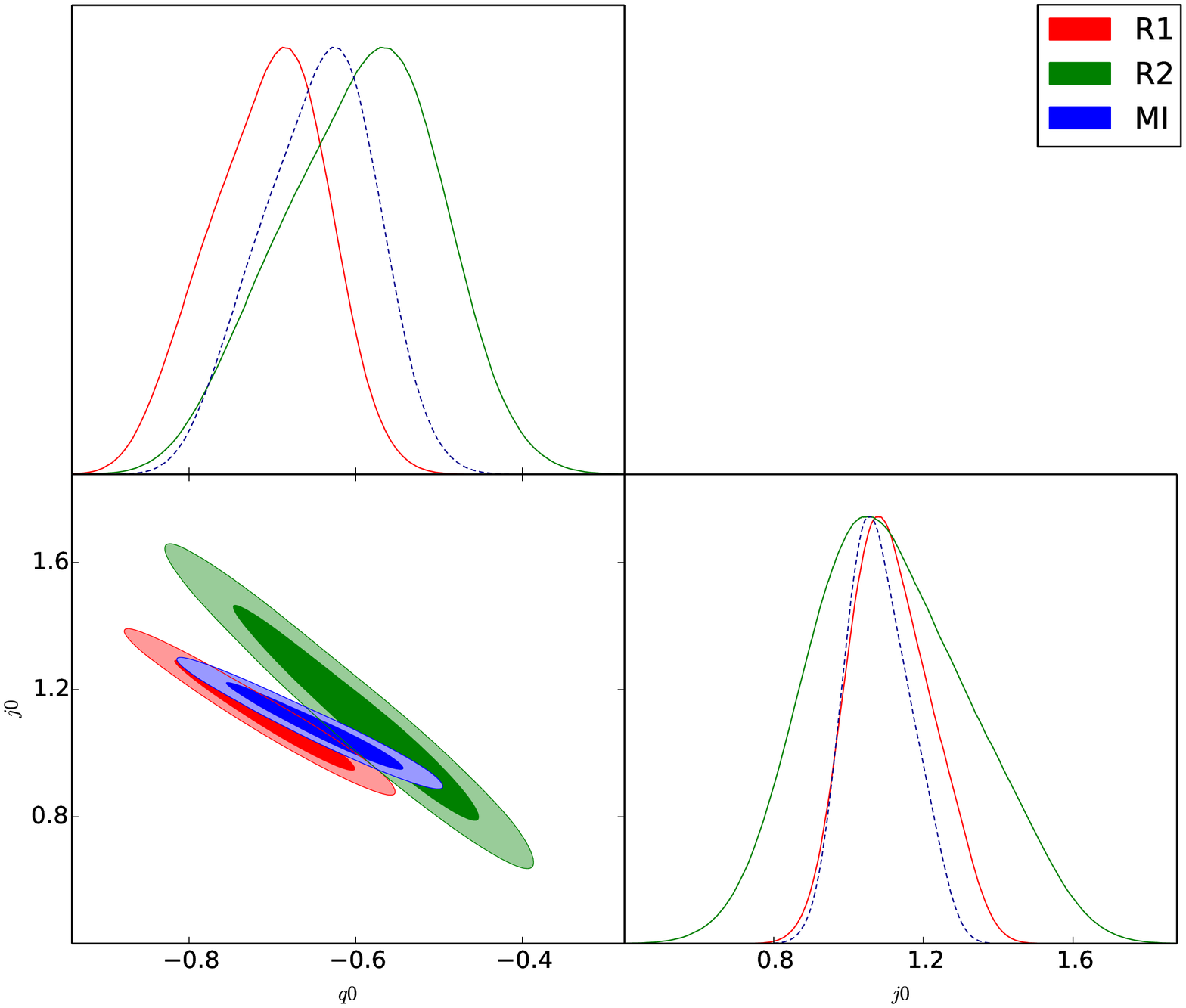}
	\includegraphics[width=8.5cm]{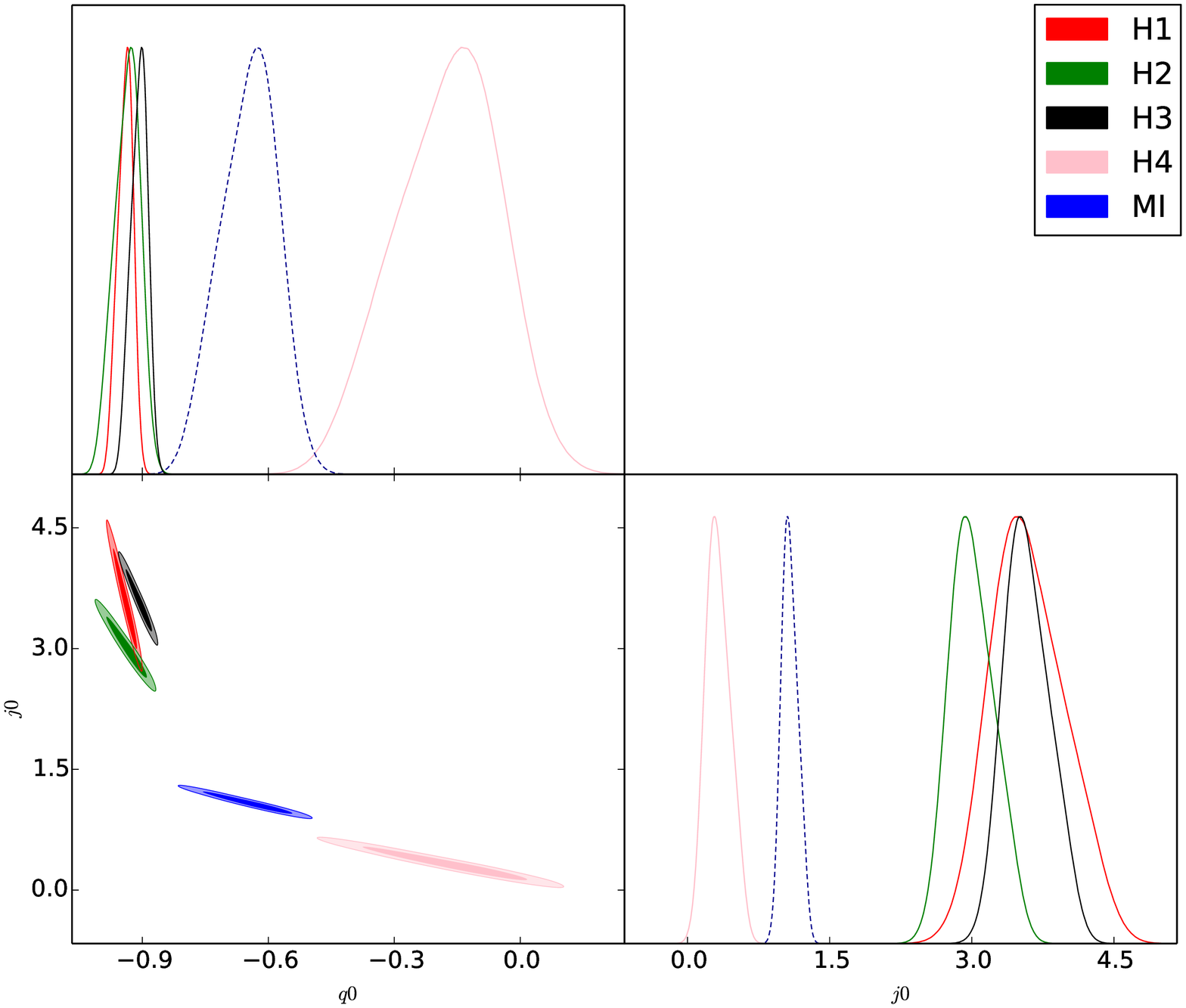}
    \includegraphics[width=8.5cm]{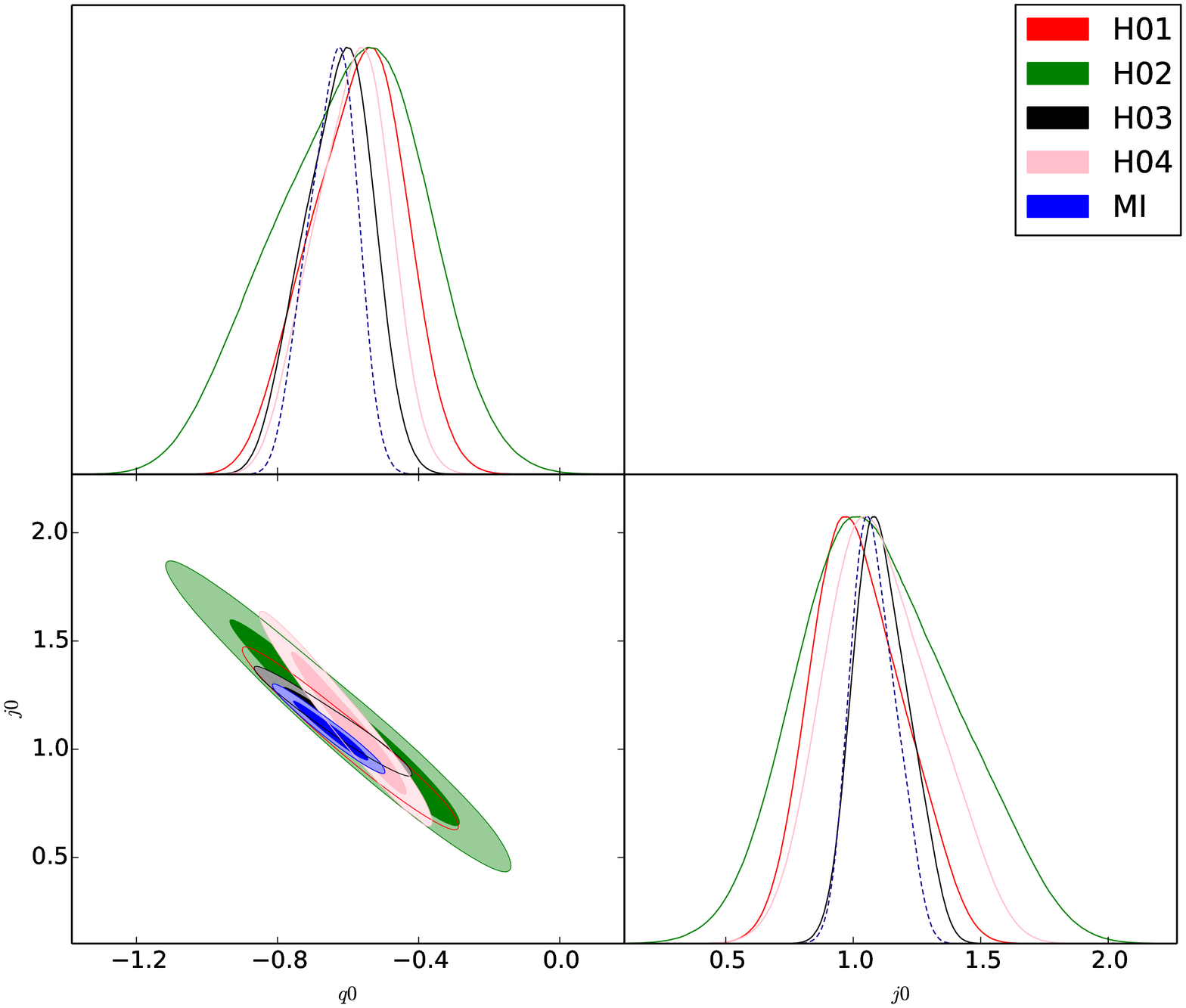}
		\caption{The results of $q_0-j_0$ space up to $2-\sigma$ confidence level obtained for different types of DE models compared with model independent results. The concordance model (top-left panel), RVMs (top-right panel), original HDE (bottom-left panel) and HDE with varriying $c^2$ term (bottom-right panel). In all the panels we plot MI result as the reference for comparison}.
	\label{fig:jplot}
\end{figure*}

Let us now perform a comparison between the cosmographic parameters of the DE Models and the cosmographic parameters obtained from the MI  approach presented in Table (\ref{tab:bestmi}).
In Fig. \ref{fig:q}, we can observe the best computed value of $q_0$ and the associated  $1\sigma$ and $2\sigma$  c.l. (confidence level) for different DE models as well as the corresponding MI results. As one can see in this figure, both running vacuum models R1, R2, as well as the holographic models  H01, H02 and H03 (based on using $L=H_0^{-1}$)  are consistent with the MI approach at $1\sigma$ c.l. In this case, however, models H04 and $\Lambda$CDM are consistent with the MI results at $2\sigma$ c.l. only. The computed value of $q_0$ for all the HDE models based on the dynamical holographic length ( $L=H^{-1}$), namely models H1, H2, H3, and H4, are in a tension with the MI results at more than $2\sigma$.  We proceed in a similar way with the jerk parameter, $j_0$, up to $2\sigma$ c.l., see Fig. \ref{fig:j}. Comparing different DE models with model independent results in this figure, we observe that H1, H2, H3, and H4 lead to manifestly  poor results. The computed values of $j_0$ for these models deviate by many standard deviations  with respect to the  MI value $j_0=1.074$, as it is plain from  Fig. \ref{fig:j}.  With a smaller tension in this case, we have model R1 which is consistent with the MI result at $2\sigma$ c.l. The other DE models, (R2,H01, H02, H03, H04 and concordance $\Lambda$CDM) are in good agreement with the model independent result at $1\sigma$ c.l.  Comparing the best fit values of $j_0$ from the different models, we find that model H02 is the most consistent one with respect to the MI result concerning the jerk parameter. This is confirmed  in Fig.\ref{fig:jplot}, where a confrontation is made with the model independent approach.
From the foregoing results we can assert  that the HDE models with dynamical holograph length are in severe disagreement with the MI results from the cosmographic parameters $q_0$ and $j_0$, and on this basis we judge that they are ruled out on  pure cosmographic grounds.  In contrast, the HDE models   using fixed holographic length equal to the current value of the horizon, viz.  $L=H_0^{-1}$, are able to pass the test since they are in agreement with the MI cosmographic approach. However, the important drawback for these HDE models  is that the option with rigid holographic length $L=H_0^{-1}$ is, of course,  the one which is theoretically less appealing.  As for the concordance model $\CC$CDM, we cannot reach a firm conclusion using the cosmographic approach since the corresponding result from $q_0$  is at the border line, whereas the  $j_0$ value for the $\CC$CDM (which is well-known to be a fixed point $j_0=1$ of the cosmological evolution\,\cite{Visser:2003vq}) it proves consistent with the MI result.

\subsection{Selection criteria}\label{sec:InformationCriteria}

After using the cosmographic approach as a procedure for model comparison, we may apply three well known methods of model selection, the Akaike information criterion (AIC) \citep{Akaike:1974}, the Bayesian information criterion (BIC) \citep{Schwarz:1974} and the deviance information criterion (DIC)\citep{Spiegelhalter:2002yvw}. These methods have been used in literature in order to compare different theoretical models with observational data.
In particular, for AIC and BIC we have:
\begin{eqnarray}
{\rm AIC} = \chi^2_{\rm min}+2k\;,\nonumber\\
{\rm BIC} = \chi^2_{\rm min}+k\ln N\;.
\end{eqnarray}
where $\chi^2_{\rm min}$ is the minimum value of $\chi^2_{\rm tot}$, $k$ is the number of model parameters and $N$ is the total number of observational data points.
The other criterion, DIC,  employs both Bayesian statistics and information theory concepts and has the following form  \citep{Spiegelhalter:2002yvw,Liddle:2007fy}:
\begin{eqnarray}\label{eq:DIC}
{\rm DIC} =\chi^2_{\rm min}+2p_D= 2\overline{\chi^2_{\rm tot}}-\chi^2_{\rm min}\;,
\end{eqnarray}
in which $p_D=\overline{\chi^2_{\rm tot}}- \chi^2_{\rm min}$  defines the effective number of parameters of the model and $2p_D$ is sometimes called the model complexity.
Overlines denote an average over the posterior distribution. Running the MCMC algorithm and using its results in the different models, we can determine the numerical values of the information criteria. In Tab. \ref{infop}, we report the numerical results of this step of our analysis for different DE models. In each column of the table, we sort out  the models according to the values of the  information criteria. We have computed $\Delta AIC$, $\Delta BIC$ and $\Delta DIC$, in which $\Delta$ means in all cases the differences between the values of the information criteria for each model, minus the corresponding  value for the best model. In the case of AIC, $\Delta AIC<2$ indicate significant support to a given model versus the best model. Furthermore, models receiving AIC within $3–7$ of the best model have considerably less support \citep{Burnham2002}. These rules of thumb appear to work also reasonably well for DIC \citep{Spiegelhalter:2002yvw}, which is perhaps the most powerful of these criteria since the numerical values is derived directly from the  the Markov chains of the  MCMC algorithm. As for BIC, the differences $\Delta BIC$ are used to gauge the evidence against a given model as compared to the best one\, \citep[for more details we refer the reader to][]{Kass:1995loi,Burnham2002,Rezaei:2019xwo,Rezaei:2021qpq}.

Furthermore, as another method of model selection in this work we have used the Bayesian evidence, or model likelihood. It comes from a full implementation of Bayesian inference at the model level and is the probability of the data given the model:
\begin{eqnarray}\label{be4}
\varepsilon \equiv p({\cal D}\mid {\cal M})=\int p(\Theta \mid {\cal M}) p({\cal D} \mid \Theta , {\cal M}) d\Theta\;,
\end{eqnarray}
 For a detailed discussion about Bayesian evidence and its use in model comparison analysis,  we refer the reader to \citep{Sola:2018sjf,Rezaei:2020mrj,Rezaei:2021qpq}.  
Given two models $M_1$ and $M_2$,  Jeffreys scale can be obtained from $\Delta\ln \varepsilon = \ln \varepsilon _{M_1}- \ln \varepsilon _{M_2}$ \citep{jeff.book}, and then the following holds:
$\Delta\ln \varepsilon < 1.1$ indicate that there is weak evidence against model $M_2$ compared with model $M_1$;  $1.1<\Delta\ln \varepsilon < 3$ indicate that there is definite evidence against model $M_2$.
Finally,  $3<\Delta\ln \varepsilon$ points to strong evidence against model $M_2$.
The various DE models under consideration are compared through these information criteria and Bayesian evidence in Table \ref{infop}. In the case of Bayesian evidence, we have assumed the  $\Lambda$CDM as the reference model and defined $\Delta\ln \varepsilon= \ln \varepsilon _{\Lambda}- \ln \varepsilon _{model}$. In the subsequent Table  \ref{chartinf} we quote the level of support (or evidence against) the given models on the basis of the numerical evaluation of the information criteria.  We mark off  the best model with boldface type.

From Tables  \ref{infop} and \ref{chartinf}  the results on comparing the various DE models from the cosmographical standpoint become  clearer.  Let us take the powerful DIC criterion, involving the sum of model fit and model complexity contributions, as indicated in Eq.\,\eqref{eq:DIC}.  It leads to a big fault for all of the  HDE models with dynamical IR cutoff $L=H^{-1}$.  This result is also confirmed by the other two information criteria, AIC \& BIC. Our results from this part show that this type of models is not favorable at all. Among the other DDEs which we have investigated, both RVM models are substantially supported, as well as models  H01 and H03 with the rigid IR cutoff $L=H_0^{-1}$, along with  the concordance $\CC$CDM model.

\begin{table*}
 \centering
 \caption{The different information criteria obtained for RVM, HDE models and the concordance $\Lambda$CDM.
}
\begin{tabular}{c  c  c c c  c c c c c c c}
\hline \hline
Model & $\chi^2_{\rm min}$ & $k$ & $N$ & $AIC$ & $BIC$& $DIC$ & $\ln \varepsilon$ & $\Delta AIC$ & $\Delta BIC$ & $\Delta DIC$ & $\Delta \ln \varepsilon$ \\
\hline
R1 & $1051.3$ & $4$ & $1083$ &  $\textbf{1059.3}$ & $1079.2$ &  $1060.0$  &$-52.44$ & $\textbf{0.0}$ & $3.6$ & $0.5$ &$-0.02$\\
\hline
R2 & $1051.5$ & $4$ & $1083$ &  $1059.5$ & $1079.4$ &  $1060.7$  &$\textbf{-52.40}$ & $0.2$ & $3.8$ & $1.2$& $\textbf{-0.06}$ \\
\hline
H1 & $1054.8$ & $4$ & $1083$ &  $1062.8$ & $1082.7$ &  $1081.1$  &$-55.29$ & $3.5$ & $7.1$ & $21.6$& $2.83$ \\
\hline
H2& $1055.0$ & $4$  & $1083$ &  $1063.0$ & $1082.9$ &  $1082.7$  & $-55.01$ &$3.7$ & $7.3$ & $23.2$& $2.55$ \\
\hline
H3 & $1054.3$ & $4$ & $1083$ & $1062.3$ & $1082.2$ &  $1082.5$  &$-55.64$ &$2.6$ & $6.6$ & $23.0$& $3.18$ \\
\hline
H4 & $1055.3$ & $4$ & $1083$ &  $1063.3$ & $1083.3$ &  $1082.8$  &$-56.14$ &$3.6$ & $7.7$ & $23.3$& $2.68$ \\
\hline
H01 & $1051.9$ & $4$ & $1083$ &  $1059.9$ & $1079.8$ &  $1061.0$  &$-53.48$ &$0.6$ & $4.2$ & $1.5$& $1.02$ \\
\hline
H02& $1051.9$ & $4$ & $1083$ &  $1059.9$ & $1079.9$ &  $1064.1$ &$-53.13$ & $0.6$ & $4.3$ & $4.6$ & $0.67$\\
\hline
H03 & $1051.9$ & $4$ & $1083$ & $1059.9$ & $1079.8$ &  $\textbf{1059.5}$& $-52.89$ & $0.6$ & $4.2$ & $\textbf{0.0}$& $0.43$ \\
\hline
H04 & $1052.6$ & $4$ & $1083$ &  $1060.6$ & $1080.6$ &  $1061.6$&$-54.70$  & $1.3$ & $5.0$ & $2.1$& $1.61$ \\
\hline
$\Lambda$CDM & $1054.6$ & $3$ & $1083$ &  $1060.6$ & $\textbf{1075.6}$ &  $1060.8$&$-52.46$  & $1.3$ & $\textbf{0.0}$ & $1.3$& $0.0$ \\
\hline \hline
\end{tabular}\label{infop}
\end{table*}

\begin{table*}
\centering
\caption{The support to (respectively, evidence against) a given DE model obtained using the results of different information criteria and Bayesian evidence (BE).
 }
 \begin{tabular}{  c  c c c  c}
 \hline \hline
Model &  AIC & BIC & DIC & BE\\
 \hline
R1 &  \bf{Best model}  & Mild to positive evidence  & \bf{Significant support} & \bf{Significant support} \\
R2 & \bf{Significant support}   &   Mild to positive evidence  & \bf{Significant support} &  \bf{Best model} \\
\hline
H1 &  Considerably less support  &   Strong evidence  & Essentially no support  & Definite evidence\\
H2 &  Considerably less support  &   Strong evidence  & Essentially no support  & Definite evidence\\
H3 &  Considerably less support  &   Strong evidence  & Essentially no support  & Strong evidence\\
H4 &  Considerably less support  &   Strong evidence  & Essentially no support  & Definite evidence\\
\hline
H01 & \bf{Significant support}  &   Mild to positive evidence  & \bf{Significant support} & Weak evidence  \\
H02 & \bf{Significant support}  &   Mild to positive evidence  & Considerably less support & Weak evidence \\
H03 & \bf{Significant support}  &   Mild to positive evidence  &  \bf{Best model} & Weak evidence \\
H04 & \bf{Significant support}  &   Mild to positive evidence  & Considerably less support& Definite evidence  \\
\hline
$\Lambda$CDM & \bf{Significant support}   &  \bf{Best model}  & \bf{Significant support} & \bf{Significant support}  \\
\hline \hline
\end{tabular}\label{chartinf}
\end{table*}

\section{Conclusions} \label{conlusion}

In this paper, using  robust cosmographic observables consisting of  the  supernovae (SnIa) data from the Pantheon sample, the  Baryonic acoustic oscillations (BAO)   from the power spectrum and bispectrum of BOSS (DR12), the complete SDSS III $Ly\alpha$-quasar  and the SDSS-IV extended BOSS DR14 quasar sample, together with the data points from cosmic chronometers (CCHs) from various sources,  we have compared different dynamical dark energy (DDE)  models  with the concordance $\Lambda$CDM model.  
In summary, our main results read as follows:

\begin{itemize}
\item The concordance $\Lambda$CDM appears to be consistent with the  cosmographic data used  in our work and passes reasonably well the tests of Bayesian evidence and the three information criteria that we have used.

\item RVM models R1 and R2 appear to be  consistent with the cosmographic data  and are also confirmed by the three  information criteria.  AIC and DIC  actually prefer the RVMs over the $\CC$CDM, whereas the  BIC test shows a mild preference for the $\CC$CDM. Bayesian evidence  shows that the RVMs are the best DE models under investigation

\item  The HDE models with dynamical holographic length $L=H^{-1}(t)$  show considerable difficulty to fit the observational data.  Despite of the fact that the alternative HDE models with fixed holographic horizon $L=H_0^{-1}$ behave better than the previous ones and can be competitive with the $\CC$CDM and the RVMs, the models with dynamical horizon $L=H^{-1}(t)$ should be the natural option for HDE.  The reason is that the holographic surface  of the universe changes with the expansion, so the past history did not have the holographic surface that we have today.  The upshot is that the natural version of the holographic models, namely those implemented with  dynamical IR cutoff $L=H^{-1}(t)$,  is not supported by the results of our analysis.  This leaves the RVMs as the only theoretically motivated  DDE models in this study which are supported by our cosmographic analysis in a competitive way with the $\CC$CDM. This does not exclude, of course, the possibility that other HDE  realizations beyond those considered here  might  be theoretically sound and phenomenologically acceptable.

\end{itemize}
Let us finally mention that the results we have obtained for the  RVMs  are in  full agreement with those from our previous studies \citep{Rezaei:2019xwo,Rezaei:2021qwd} which used different data samples, including non-cosmographic data.  Overall the RVMs  appear as phenomenologically consistent models which can compete with the concordance cosmology   and have also the capacity to alleviate the current tensions \,\citep{Rezaei:2019xwo,Sola:2021txs}.

\section{Acknowledgements}
JSP acknowledges partial support by  projects  PID2019-105614GB-C21 and FPA2016-76005-C2-1-P (MINECO, Spain), 2017-SGR-929 (Generalitat de Catalunya) and CEX2019-000918-M (ICCUB). JSP also acknowledges participation in the COST Association Action CA18108 ``{\it Quantum Gravity Phenomenology in the Multimessenger Approach (QG-MM)}''.

\section{appendix: EoS for the DE models}
Adopting the cosmological parameters \eqref{omega}, one can write Eq.\eqref{frid1} in terms of the well-known cosmic sum rule
\begin{eqnarray}\label{omegasum}
\Omega_{\rm r}+ \Omega_{\rm m}+\Omega_{\rm d}=1\,.
\end{eqnarray}
Furthermore, the conservation equations for the corresponding energy densities are as follows:
 \begin{eqnarray}\label{continuity}
 && \dot{\rho_{\rm r}}+4H\rho_{\rm r}=0\,,\label{radiation}\\
&&\dot{\rho_{\rm m}}+3H\rho_{\rm m}=0\,,\label{matter}\\
&&\dot{\rho_{\rm d}}+3H(1+w_{\rm d})\rho_{\rm d}=0\,,\label{de},
 \end{eqnarray}
where the over-dot stands for derivative with respect to cosmic time $t$.  We emphasize here that the DE is assumed to be self-conserved without interaction with matter in the current approach, as it is obvious from the previous equations.  The above equations are valid both for the RVMs and the HDE's considered here.

As indicated in the text, the EoS for the RVMs is given by Eq.\,\eqref{eq:EoSmod3} and details have been given in  \citep{Rezaei:2019xwo}. Therefore, in the following we focus on the calculation of the EoS for the various HDEs under consideration.
Assuming first that the current value of the inverse  Hubble parameter as the effective  IR cutoff in Eq.\eqref{hde}, i.e. $L=H_0^{-1}$,  the energy density for the different types of HDE models take the following form:
\begin{eqnarray}\label{GHDE}
\rho_{\rm d}(z)=3 c^2(z) M^2_{\rm p}H_0^2\;.
\end{eqnarray}
The HDE energy density  after using equations.\eqref{omega} and \eqref{GHDE} take on the form
\begin{eqnarray}\label{omeghde}
\Omega_{\rm d}=\frac{c^2(z)}{E^2(z)}\;.
\end{eqnarray}
where $E(z)=H(z)/H_0$. Using the functions of $\Omega_{\rm i}$, and replacing Eq.\eqref{omeghde} into Eq.\eqref{omegasum} we have:
\begin{eqnarray}\label{e20}
\frac{\Omega^0_{\rm m}(1+z)^3}{E^2(z)}+\frac{\Omega^0_{\rm r}(1+z)^4}{E^2(z)}+\frac{c^2(z)}{E^2(z)}=1\;,
\end{eqnarray}
where $\Omega^0_{\rm m}$ and $\Omega^0_{\rm r}$ are the current values of the matter and radiation density parameters respectively. 
Therefore the normalized Hubble parameter, $E(z)$ takes the following form
\begin{eqnarray}\label{e2}
E(z)=\sqrt{\Omega^0_{\rm m}(1+z)^3+\Omega^0_{\rm r}(1+z)^4+c^2(z)}\;.
\end{eqnarray}
If we take the time derivative of Eq.\eqref{GHDE}, we find
\begin{eqnarray}\label{GHDEdot}
\dot{\rho}_{\rm d}=2\rho_{\rm d}\frac{\dot{c}(z)}{c(z)}\;.
\end{eqnarray}
The EoS parameter of the HDE models is obtained by inserting  Eq.\eqref{GHDEdot} into Eq.\eqref{de}:
\begin{eqnarray}\label{EOS}
w_{\rm d}=-1-\frac{2}{3}\frac{c'(z)}{c(z)}\;,
\end{eqnarray}
where prime stands for derivative with respect to the variable $x\equiv\ln a=-\ln (1+z)$, i.e.
\begin{equation}\label{eq:primederiv}
  c'(z)=\frac{dc(z)}{dx}=-\frac{dc(z)}{d\ln(1+z)}=-(1+z)\frac{dc(z)}{d z}\,.
\end{equation}
Clearly, for  $c$=const., we have $c'(z)=0$, which leads to $w_{\rm d}=-1$ (the concordance $\Lambda$-cosmology).
In order to find the cosmographic parameters in HDE cosmology, from Eq.\eqref{eq2} we obtain:
\begin{eqnarray}
q=-\dfrac{a\ddot{a}}{\dot{a}^2}=-1-\dfrac{\dot{H}}{H^2}\;.\label{qqq}
 \end{eqnarray}
Furthermore, following the standard lines,  it is easy to find \citep{Rezaei:2017yyj}
\begin{eqnarray}
\dfrac{\dot{H}}{H^2}=-\dfrac{3}{2}(1+w_{\rm d}\Omega_{\rm d})\;.\label{qq}
\end{eqnarray}
Inserting Eqs.(\eqref{omeghde} and \eqref{EOS}) in Eq.\eqref{qq} and replacing its result in Eq.\eqref{qqq} we can derive the deceleration parameter $q$ in HDE models:
\begin{eqnarray}\label{eq:computedq0}
q=\dfrac{1}{2}-\dfrac{3}{2}\dfrac{c^2(z)}{E^2(z)}-\dfrac{c'(z)c(z)}{E^2(z)}\;.\label{q}
\end{eqnarray}
where $c'(z)$ is the first derivative of $c(z)$ with respect to $\ln a$. For the other cosmographic parameters, using Eqs.(\eqref{eq1}-\eqref{eq5}) and Eq.\eqref{q} we arrive at \citep{Aviles:2012ir}:
\begin{eqnarray}\label{eq:computedj0}
j=-\dfrac{dq}{d \ln a}+2(1+q)^2-3q-2\,,\label{j}
\end{eqnarray}
where $j$ is the jerk parameter. Furthermore, in the case of the snap $s$ and the lerk parameter $l$, we have:
\begin{eqnarray}
s=\dfrac{dj}{d \ln a}-2j-3qj\,,\label{s}
\end{eqnarray}
\begin{eqnarray}
l=\dfrac{ds}{d \ln a}-3s-4qs\,.\label{l}
\end{eqnarray}
Now by choosing the various ansatz functions for $c(z)$ indicated  in Eqs.(\eqref{par1}-\eqref{par4})  we  may derive the current values of the cosmographic parameters $(q_0, j_0, s_0, l_0)$  for each of the HDE cosmologies under study. They are obtained from the above formulas evaluated at $z=0$.

As another choice for the effective  IR cutoff in Eq.\eqref{hde}, we have  the dynamical holographic length  $L=H^{-1}(t)$.  In this case, the energy density for the different types of HDE models adopt the following form:
\begin{eqnarray}\label{GHDE1}
\rho_{\rm d}(z)=3 c^2(z) M^2_{\rm p}H^2\;.
\end{eqnarray}
As for the HDE energy density, it follows  after using equations.\eqref{omega} and \eqref{GHDE1}:
\begin{eqnarray}\label{omeghde1}
\Omega_{\rm d}=c^2(z)\;.
\end{eqnarray}
Using the functions of $\Omega_{\rm i}$, and replacing Eq.\eqref{omeghde1} into Eq.\eqref{omegasum},  we have:
\begin{eqnarray}\label{e201}
\frac{\Omega^0_{\rm m}(1+z)^3}{E^2(z)}+\frac{\Omega^0_{\rm r}(1+z)^4}{E^2(z)}+c^2(z)=1\;,
\end{eqnarray}
The Hubble parameter in this case reads
\begin{eqnarray}\label{e21}
E(z)=\sqrt{\frac{\Omega^0_{\rm m}(1+z)^3+\Omega^0_{\rm r}(1+z)^4}{1-c^2(z)}}\;.
\end{eqnarray}
On taking the time derivative of Eq.\eqref{GHDE1}, we find
\begin{eqnarray}\label{GHDEdot1}
\dot{\rho}_{\rm d}=2\rho_{\rm d}\left(\frac{\dot{c}(z)}{c(z)}+\frac{\dot{H}(z)}{H(z)}\right)\;.
\end{eqnarray}

In order to find the EoS parameter of HDE models we insert Eq.\eqref{GHDEdot1} into Eq.\eqref{de}:
\begin{eqnarray}\label{EOS1}
w_{\rm d}=-1-\frac{2}{3}\left(\frac{c'(z)}{c(z)}+\frac{H'(z)}{H(z)}\right)\;,
\end{eqnarray}
where prime still means derivative with respect to the variable $x\equiv\ln a=-\ln (1+z)$.
Following  similar steps as in the previous case, one can easily find out  the cosmographic parameters for the HDE cosmology with dynamical $L=H^{-1}(t)$, and we just omit details.


\subsection{The CPL parametrization, H1}
 The EoS parameter for these models may now be found straightforwardly. Using Eq.\eqref{par1},we find $c \rightarrow c_0+c_1$ for the asymptotic value in the past $(z\rightarrow\infty)$, whereas at the present time $z\rightarrow 0$  we have $c\rightarrow c_0$. Therefore, in this case, $c$ varies slowly from $c_0+c_1$ to $c_0$.   From equations \eqref{EOS} and \eqref{eq:primederiv} we find  immediately the EoS parameter for model H1:
\begin{eqnarray}\label{EOS1}
w_{\rm d}=-1+\frac{2}{3}\frac{c_1}{c_0(1+z)+c_1z}\;,
\end{eqnarray}
We can see that  $w_{\rm d}(z=0)=-1+\frac23\frac{c_1}{c_0}$  and $ w_{\rm d}(z\to\infty)=-1$ (only at this asymptotic point in the past behaves like vacuum).  Obviously the fact that any EoS must be close to the vacuum one near our time (as suggested by observations\,\citep{Aghanim:2018eyx}) enforces $|c_1|\ll c_0$ for this model.
\subsection{The JBP parametrization, H2}
Similarly,  equations  \eqref{par2}  and  \eqref{EOS}  render the EoS for H2:
\begin{eqnarray}\label{EOS2}
w_{\rm d}=-1+\frac{2}{3}\frac{c_1(1-z)}{c_0(1+z)^2+c_1z}\,.
\end{eqnarray}
It is easy to see from \eqref{par2}  that both at $z=0$ and  for $z\rightarrow\infty$ we have $c=c_0$.  However, the EoS values at these points are different and coincide with the previous model: $w_{\rm d}(z=0)=-1+\frac23\frac{c_1}{c_0}$  and
$ w_{\rm d}(z\to\infty)=-1$. Once more consistency of the  EoS  with measurements leads to $|c_1|\ll c_0$ also for this model.

\subsection{The Wetterich parametrization, H3}
In  Wetterich-type, Eq. \eqref{par3}, at $z=0$, we have $c=c_0$ while in the asymptotic past,  $z\rightarrow\infty$,  we have $c=0$. Thus Eq. \eqref{hde} tells us that we can ignore the role of the HDE for this model at the early times.  Calculation yields
\begin{eqnarray}\label{EOS3}
w_{\rm d}=\frac{2}{3}\frac{c_1\left( 1+c_1\ln(1+z)\right) }{c^2_0-\left( 1+c_1\ln(1+z)\right)^2}\,.
\end{eqnarray}
At the present time we have  $w_{\rm d}(z=0)=\frac23\frac{c_1}{c_0^2-1}$ and this enforces the peculiar (fine-tuned) relation  $c_0^2\simeq 1 -\frac23 c_1$,   as well as  $0<c_1<3/2$.   The lower limit  $c_1>0$ must hold  as otherwise the previous relation implies $c_0>1$, in contradistinction to Eq.\,\eqref{coOmegaLambda}. This entails that $w_{\rm d}<0$ at all points in the past while it approaches $w_{\rm d}=0$ for $z\to\infty$.  There is also a point where there is a singular behavior, which is when the denominator vanishes and thus $w_{\rm d}\to\infty$.   This point occurs at
\begin{equation}\label{eq:singularpoint}
\ln(1+z)=\frac{c_0-1}{c_1}<0
\end{equation}
which brings the singularity at a point in the future ($-1<z<0$).

\subsection{The Ma-Zhang parametrization, H4}

For the last form of $c(z)$, the Ma-Zhang parametrization, Eq.\eqref{par4}, we observe that at $z=0$ we have $c(z)=c_0$, while at $z\rightarrow\infty$, we have $c(z)= c_0 -c_1\ln 2$. It is easy to see that in the far future, when $z \rightarrow-1$, $c(z)$ diverges.  Taking the derivative $d/dx$  of Eq.\eqref{par4} we find

\begin{eqnarray}\label{par4prim}
c'(z)=\frac{c_1\ln(2+z)}{(1+z)}-\frac{c_1}{(2+z)}\;.
\end{eqnarray}
Therefore,  Eq. \eqref{EOS} leads to
\begin{eqnarray}\label{EOS4}
&w_{\rm d}=\frac{2}{3}\frac{c_1(1+z)^2\left[ (1+z)+(2+z)\ln(2+z)\right]}{(2+z)\left[c_0(1+z)+c_1\ln(2+z)-c_1(1+z)\ln2 \right]}\nonumber \\
&\times\frac{1}{(1+z)^2-\left(c_0(1+z)+c_1\ln(2+z)-c_1(1+z)\ln2\right)^2}\,.
\end{eqnarray}
For $z\to\infty$,  $w_{\rm d}\to 0$, so it behaves as dust.  There are singular points  where  the denominator vanishes and as a consequence  $w_{\rm d}\to\infty$. It can be seen that this may happen at a point $z$ satisfying $(c_0-c_1\ln 2) (1+z)=-c_1\ln(2+z)$ and  at a point satisfying $(1-c_0+c_1\ln 2) (1+z)=c_1\ln(2+z)$ .   Since $|c_1|\ll c_0$ these  points can exist only in the future, the first exists  if $c_1<0$ and the second  if $c_1>0$. There are no EoS singularities in the past.
In contrast, for the current epoch ($z=0$) we expect that $w_{\rm d}(0)\simeq -1$ and this enforces the following approximate relation between the parameters
\begin{equation}\label{eq:relationparam}
\frac{c_1}{c_0(1-c_0^2)}\simeq -\frac{3}{1+2\ln 2}\,.
\end{equation}
From the foregoing and Eq.\,\eqref{coOmegaLambda}  it follows that this relation can only be satisfied for $c_1<0$. It follows that of the aforementioned singular points only the first exists.

\bibliographystyle{apsrev4-1}

\end{document}